\documentclass[final,english]{bullsrsl}[2022/06/15]
\usepackage[latin1]{inputenc}
\usepackage[T1]{fontenc}
\usepackage{graphicx}	
\usepackage{amsmath}	
\usepackage{amssymb}
\usepackage{subfigure}
\usepackage{amsfonts}
\usepackage{natbib}
\usepackage{hyperref}
\usepackage{booktabs}
\hypersetup{
    colorlinks=true,
    linkcolor=blue,
    filecolor=blue,      
    urlcolor=red,
    citecolor=blue,
}

\newcommand{\po}{P$_{\Omega}$\,} 
\newcommand{\ps}{P$_{\omega}$\,} 
\newcommand{\ptwos}{P$_{2\omega}$\,} 
\newcommand{\pb}{P$_{\omega-\Omega}$\,} 
\newcommand{\ptwob}{P$_{2(\omega-\Omega)}$\,} 

\begin{document}
\title{Timing Analysis of the Intermediate Polar - V709 Cas}

\author[affil={1}, corresponding]{Srinivas M}{Rao}
\author[affil={1}]{Jeewan Chandra}{Pandey}
\author[affil={1}]{Nikita}{Rawat}
\author[affil={2}]{Arti}{Joshi}
\affiliation[1]{Aryabhatta Research Institute of observational sciencES (ARIES), Nainital 263001, India}
\affiliation[2]{Pontificia Universidad Cat\'{o}lica de Chile, Av. Vicu\~{n}a Mackenna 4860, 782-0436 Macul, Santiago, Chile}
\correspondance{srinivas22546@gmail.com}
\maketitle

\begin{abstract}
We have carried detailed time-resolved timing analysis of an intermediate polar V709 Cas using the long-baseline, short-cadence optical photometric data from the Transiting Exoplanet Survey Satellite. We found an orbital period of $5.3341\pm0.0004$ hr, a spin period of $312.75\pm0.02$ sec, and a beat period of $317.93\pm0.03$ sec, which is similar to the earlier published results. From the continuous high cadence data, we found that  V709 Cas is a disc overflow system with disc-fed dominance.
\end{abstract}


\section{Introduction}
Cataclysmic variable stars (CVs) are binary systems with a primary white dwarf (WD)  and a secondary late-type main sequence star that fills its Roche lobe in which primary accretes material from the secondary onto the primary. Magnetic and non-magnetic  CVs are two categories of CVs based on the magnetic field strength of the primary white dwarf. The WDs of non-magnetic CVs have relatively weak magnetic fields that do not affect the accretion process. In contrast, the  WD in Magnetic CVs (MCVs) have a high magnetic field and are categorized into intermediate polars (IPs) and polars. Polars have strong magnetic fields (>10 MG) and accretion flow channels along the field lines. In the case of intermediate polars, the magnetic field of the white dwarf is moderately strong, typically less than tens of MG. 

Due to the intricate interplay between the magnetic field and the accretion flow, IPs provide a unique laboratory for studying accretion processes in the presence of strong magnetic fields. Thus the accretion flow in IPs is characterized by complex and variable behaviour. Three different accretion mechanisms are found in the IPs. In disc-fed accretion, material from a Keplerian disc accretes onto the magnetic poles of the white dwarf after disruption at the magnetosphere \citep{Hellier1989a} and the accretion occurs via accretion curtains. In stream-fed accretion, material directly accretes onto the white dwarf surface along its magnetic field lines \citep{Rosen1988}. In disc overflow accretion, material overflows from the disc and interacts with the magnetosphere, resulting in both disc-fed and stream-fed accretion simultaneously \citep{Hellier1989b}. Another potential mechanism is diamagnetic blob accretion (\citealt{King1993}, \citealt{Lubow1989}). In disc-fed and disc overflow accretion, material accretes onto both magnetic poles. 

Generally, the accretion mode can be determined from periodicities in time series data. Disc-fed accretion shows spin ($\omega$) modulation, while stream-fed accretion shows predominant modulation in beat frequency ($\omega - \Omega$) \citep{Ferrario1999}, where $\Omega$ is the orbital frequency. Disc overflow accretion shows both spin and beat modulation (see \citealt{Hellier1991}, \citeyear{Hellier1993}). In this paper, we used high cadence long baseline TESS data to explore the accretion mechanism in the IP V709 Cas.

V709 Cas was recognized as intermediate polar by \cite{HaberlMotch1995} using the data obtained from ROSAT. The spin and orbital periods of V709 Cas were found to be 312.78 s and 5.33278 hr, respectively \citep{Norton1999,Kozhevnikov2001,Thorstensen2010}. \cite{Kozhevnikov2001} also found that the system modulates at the beat period of 317.94 s. It was found to be a disc-fed accreator by \cite{Norton1999} using the X-ray data obtained from ROSAT-HRI.

\section{Observations and Data Analysis}
\begin{figure}[ht]
\centering
\includegraphics[width=\textwidth]{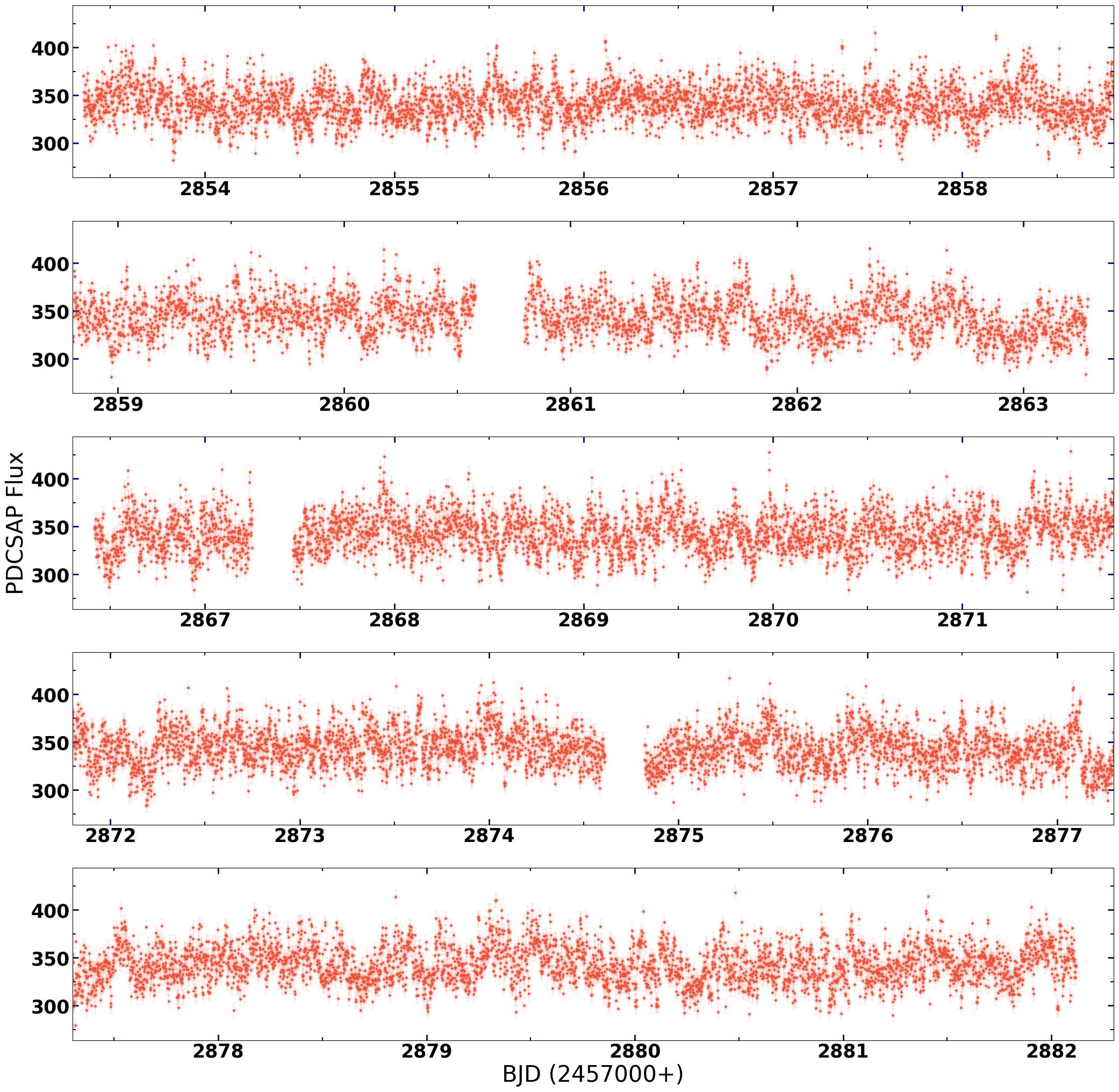}
\caption{Light curve of V709 Cas as observed from TESS during the observation in Sector 57.}
\label{lightcurve}
\end{figure}
We use the archival data from high cadence long-baseline Transiting Exoplanet Survey Satellite (TESS) observations. We refer \cite{2015JATIS...1a4003R} for detail about the TESS and related instruments.  
Briefly, TESS observes the sky in the 600 -- 1000 nm waveband with an effective wavelength of 800 nm for continuous 27.4 days. The data was stored in the Mikulski Archive for Space Telescopes data with the unique identification number 'TIC 320180973'. We have considered the PDCSAP flux data for those with the '0' quality flag. TESS observed V709 Cas in five sectors 17, 18, 24, 57 and 58. The data from all observed sectors are available in 2 minutes cadence, whereas the data from two sectors, 57 and 58, are available in a 20-s cadence. Therefore, for our preliminary analysis, we have used the data from sector 57. Due to the presence of short cadence data, we have considered sector 57 for our preliminary analysis. 

\section{Period Analysis}
\begin{figure}
    \subfigure[120 s Cadence]{\includegraphics[width=0.49\textwidth]{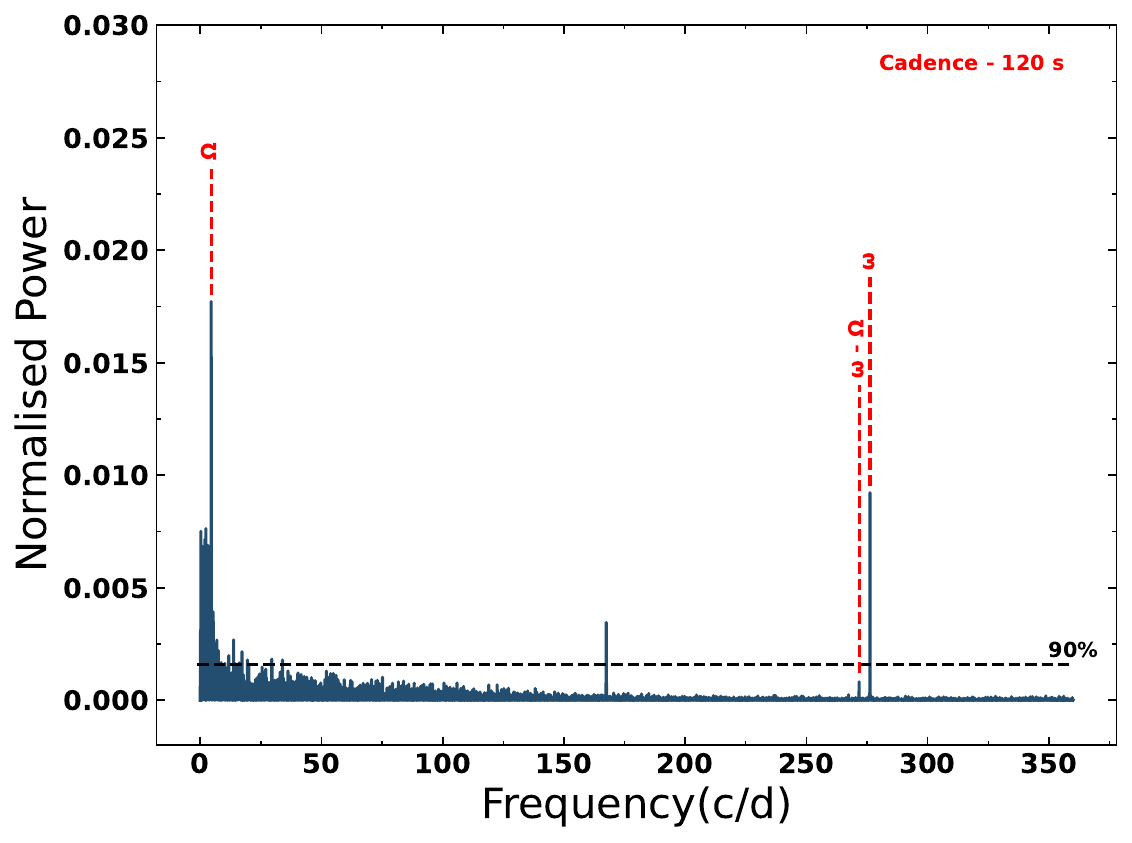}}
    \subfigure[20 s Cadence]{\includegraphics[width=0.49\textwidth]{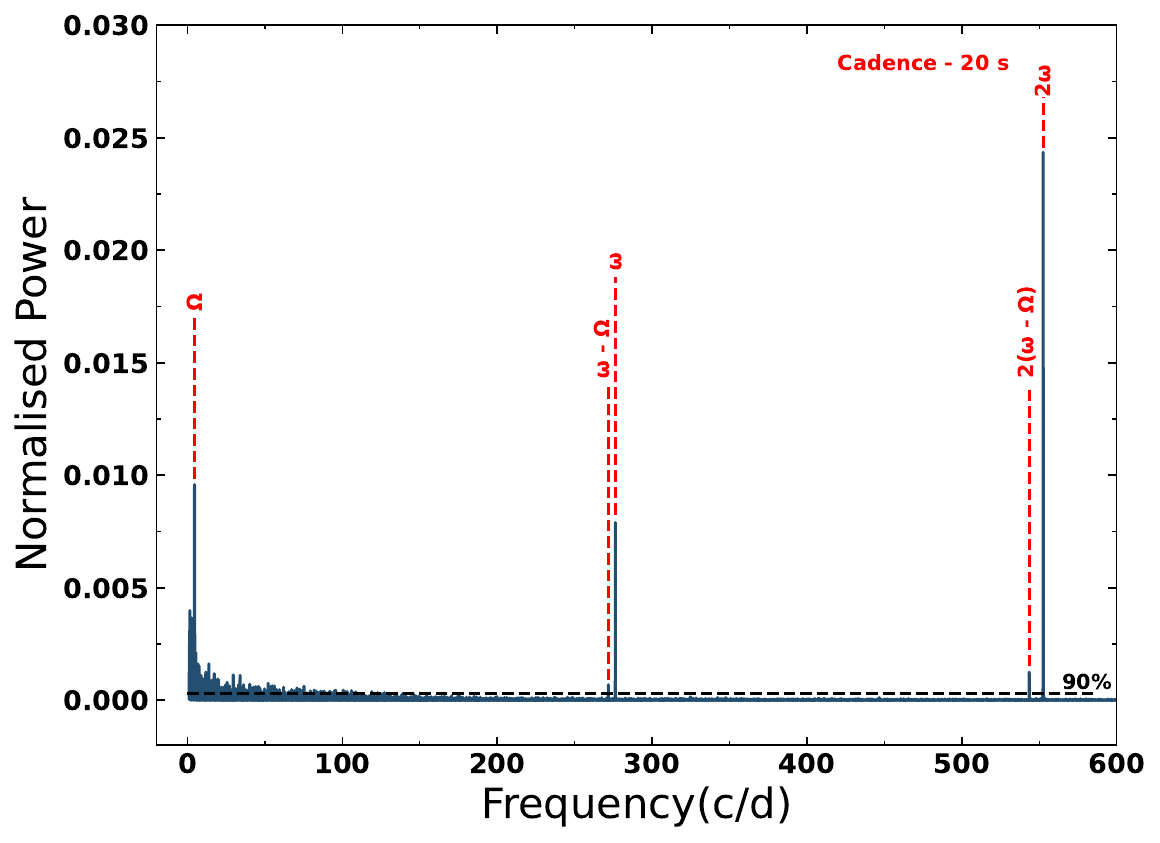}}
    \caption{Power Spectra of V709 Cas. All the periods are marked by vertical dashed lines and the FAP level is marked by the horizontal black dashed line}
    \label{power spectra}
\end{figure}
Figure \ref{lightcurve} shows the lightcurve of V709 Cas for sector 57. We have done a time-series analysis of this long and continuous optical data. The periodogram analysis was performed using the Lomb-Scargle (\citealt{Lomb1976}; \citealt{Scargle1982}) method.  The power spectra for the 120 s and 20 s cadence data are shown in Figure \ref{power spectra} (a) and Figure \ref{power spectra} (b), respectively. To ensure the presence of periods, we have set the significant level at 90$\%$ by calculating the false alarm probability (FAP) (refer \citealt{Vanderplas2018}). The FAP level is marked as a horizontal black dashed line in this Figure. For 120 s cadence data, we have detected the orbital and the spin frequencies above the FAP level. Although the beat frequency is present, but it is below the FAP level. Due to the increased Nyquist frequency limit in short cadence data, we found harmonics of spin and beat frequencies. In this case, the beat frequency and its harmonic were above the FAP level. The power of the first harmonic of spin frequency was more than that of the fundamental spin frequency. 

In Figure \ref{power spectra} for the 120 s cadence data, one can notice a peak corresponding to a frequency of $\sim$168 c/d, but no such peak is seen for the 20 s cadence data. This frequency obtained from the 120 s cadence data is due to the super-Nyquist frequency ($N_\nu$) and corresponds to 2($N_\nu - \omega$), thus not being considered an actual frequency. The periods corresponding to these significant frequencies are given in Table \ref{periods from power spectra}. 

\begin{table}
    \centering
    \caption{Periods obtained from the LS power spectra of sector 57.}
    \setlength\tabcolsep{2.5pt}
    \begin{tabular}{cccccc}
    \toprule
     \textbf{Cadence (sec)} & \textbf{\po (hr)} & \textbf{\ps (sec)} & \textbf{\pb (sec)} & \textbf{\ptwos (sec)} & \textbf{\ptwob (sec)} \\
     \midrule
    120 & $5.3341 \pm 0.0007$ & $312.74 \pm 0.04$ & $317.93 \pm 0.04$ & - & -\\
    20 & $5.3344 \pm 0.0004$ & $312.75 \pm 0.02$ & $317.93 \pm 0.03$ & $156.37 \pm 0.01$ & $158.96 \pm 0.01$\\
    \bottomrule
    \end{tabular}
    \label{periods from power spectra}
\end{table}

\section{Discussion and Conclusion}
Through Lomb-Scargle periodogram analysis of data with 120 s and 20 s cadence, we obtained similar orbital, spin, and beat periods as shown in Table \ref{periods from power spectra}. The obtained orbital periods are consistent with the earlier finding by results from \cite{Motch1996}, \cite{Bonnet-Bidaud2001}, and  \cite{Thorstensen2010}. The spin and beat periods also agree with earlier results from \cite{Norton1999} and \cite{Kozhevnikov2001}. However, we did not observe the variability reported by \cite{Hric2014}.

The power spectra show signatures of both the spin frequency ($\omega$),  beat frequency ($\omega - \Omega$) and their harmonics. The accretion mechanisms are explained based on the frequencies present in the power spectra as per \cite{Ferrario1999} and \cite{Wynn1992}. Stream-fed accretion tends to show more power at the beat frequency, while the dominant power in spin frequency indicates disc-fed accretion. Both frequencies in the power spectra indicate that  V709 Cas is a disc overflow system where part of the accretion stream bypasses the disc and directly interacts with the magnetosphere, resulting in both disc-fed and stream-fed accretion simultaneously. However, since the spin frequency is more dominant, more matter is expected to accrete via the disc rather than the stream. Therefore the accretion geometry of the system can be attributed to a disc-overflow accretion with disc-fed dominance.
\begin{acknowledgments}
This work includes data collected with the \textit{TESS} mission, obtained from the MAST data archive at the Space Telescope Science Institute (STScI). Funding for the \textit{TESS} mission is provided by the NASA Explorer Program. STScI is operated by the
Association of Universities for Research in Astronomy, Inc.,
under NASA contract NAS 5-26555.
\end{acknowledgments}

\begin{furtherinformation}

\begin{orcids}
\orcid{0009-0002-6282-8164} {Srinivas M} {Rao}
\orcid{0000-0002-4331-1867} {J. C.} {Pandey}
\orcid{0000-0002-4633-6832} {Nikita} {Rawat}
\end{orcids}

\begin{authorcontributions}
All authors contributed significantly to the work presented in this paper.

\end{authorcontributions}

\begin{conflictsofinterest}
The authors declare no conflict of interest.
\end{conflictsofinterest}

\end{furtherinformation}

\bibliographystyle{bullsrsl-en}

\bibliography{main}

\end{document}